\documentclass[aps,prl,twocolumn,showpacs,superscriptaddress,10pt]{revtex4-1}

\usepackage{amsmath}
\usepackage{amsfonts}
\usepackage{amssymb}
\usepackage{graphicx}
\usepackage{color}
\usepackage{soul}
\usepackage{xcolor}
\usepackage{braket}

\newcommand{\pvec}[1]{\vec{#1}\mkern2mu\vphantom{#1}}

\begin{document}
\title{Shattered Time: Can a Dissipative Time Crystal Survive Many-Body Correlations?}

\author{K. Tucker}
\affiliation{JILA, NIST, Department of Physics, University of Colorado,  Boulder, CO 80309, USA}
\affiliation{Department of Applied Mathematics, University of Colorado, Boulder, CO 80309, USA}
\author{B. Zhu}
\affiliation{JILA, NIST, Department of Physics, University of Colorado,  Boulder, CO 80309, USA}
\affiliation{Center for Theory of Quantum Matter, University of Colorado, Boulder, CO 80309, USA}
\affiliation{ITAMP, Harvard-Smithsonian Center for Astrophysics, Cambridge, MA 02138, USA}
\author{R. J. Lewis-Swan}
\affiliation{JILA, NIST, Department of Physics, University of Colorado,  Boulder, CO 80309, USA}
\affiliation{Center for Theory of Quantum Matter, University of Colorado, Boulder, CO 80309, USA}
\author{J. Marino}
\affiliation{JILA, NIST, Department of Physics, University of Colorado,  Boulder, CO 80309, USA}
\affiliation{Center for Theory of Quantum Matter, University of Colorado, Boulder, CO 80309, USA}
\affiliation{Department of Physics, Harvard University, Cambridge MA, 02138, USA}
\author{F. Jimenez}
\affiliation{Department of Applied Mathematics, University of Colorado, Boulder, CO 80309, USA}
\author{J. G. Restrepo}
\affiliation{Department of Applied Mathematics, University of Colorado, Boulder, CO 80309, USA}
\author{A. M. Rey}
\affiliation{JILA, NIST, Department of Physics, University of Colorado,  Boulder, CO 80309, USA}
\affiliation{Center for Theory of Quantum Matter, University of Colorado, Boulder, CO 80309, USA}
\begin{abstract}
We investigate the emergence of a  time crystal  in a driven-dissipative many-body  spin array. In this system the interplay  between incoherent
spin pumping and collective  emission stabilizes  a  synchronized non-equilibrium steady state which in the thermodynamic limit features a self-generated
time-periodic pattern  imposed by  collective elastic interactions. In contrast to  prior realizations where the time symmetry is already  broken by an external drive,
here it is only spontaneously broken  by the elastic exchange interactions and manifest in  the two-time correlation spectrum. Employing a combination of exact
numerical calculations and a second-order cumulant expansion, we  investigate the impact of many-body correlations on the  time crystal formation and   establish a connection between the regime where it is stable  and  a slow  growth rate of the mutual information, signalling that the time crystal studied here is an emergent semi-classical out-of-equilibrium state of matter. We also confirm the rigidity of the time crystal to single-particle dephasing. Finally, we discuss  an experimental implementation using long-lived dipoles in an optical cavity.
\end{abstract}

\date{\today}
\maketitle

\emph{Introduction ---}
Experimental progress in the control and preparation of quantum cold gases~\cite{Bloch2008} has opened a new era in which non-equilibrium phenomena have a central role.
In particular,  time crystals (TCs)~\cite{Wilczek, saph,Noz, Sacha, laz17,Else, Khe, Khe2, Key, Key2, Yao,yao2,Gong,Autti} -- phases of quantum matter which spontaneously break time translational invariance  and which can only exist in
out-of-equilibrium systems~\cite{Wantabe} -- have recently attracted significant attention.
A  system hosting a crystalline time phase should be many-body and exhibit an order parameter, $\phi(\vec{r},t)$, whose unequal time correlation function
approaches, in the thermodynamic limit, a non-trivial periodic, oscillating function of time~\cite{Wantabe}: $\langle \phi(\vec{r},t)\phi(\pvec{r}',0)\rangle \to f(t)$, at sufficiently large distances $\vert \vec{r} - \pvec{r}' \vert$. Such behavior must be robust to imperfections  of the system parameters or external perturbations.

So far TCs have been experimentally realized \cite{Monroe,Lukin} in periodically driven, interacting quantum many-body systems with spatial disorder, also known as
Floquet time crystals~\cite{Else, Khe, Khe2, Key, Key2, Yao}. Typically, disorder provokes the onset of a many-body localized phase or a pre-thermal state~\cite{Bruno3, aba, Wantabe, Else, Ho, ElsePRX, Kuwa, Rovny} where
heating towards infinite temperature is suppressed~\cite{ponte, ponte2, zhang, bordia}, and time-resolved observables can react to the  periodic drive with a
dynamical entrainment at a  frequency which is a subharmonic of the one imposed by the external drive. Besides those implementations, which directly break discrete time symmetry, there are other proposals of TCs that break continuous time symmetry \cite{Sacha2,Syrwid17,Nakatsugawa}. However, to our knowledge, there has been no experimental observation of a continuous TC in a quantum many-body system.

\begin{figure}[t!]
\includegraphics[width=8cm]{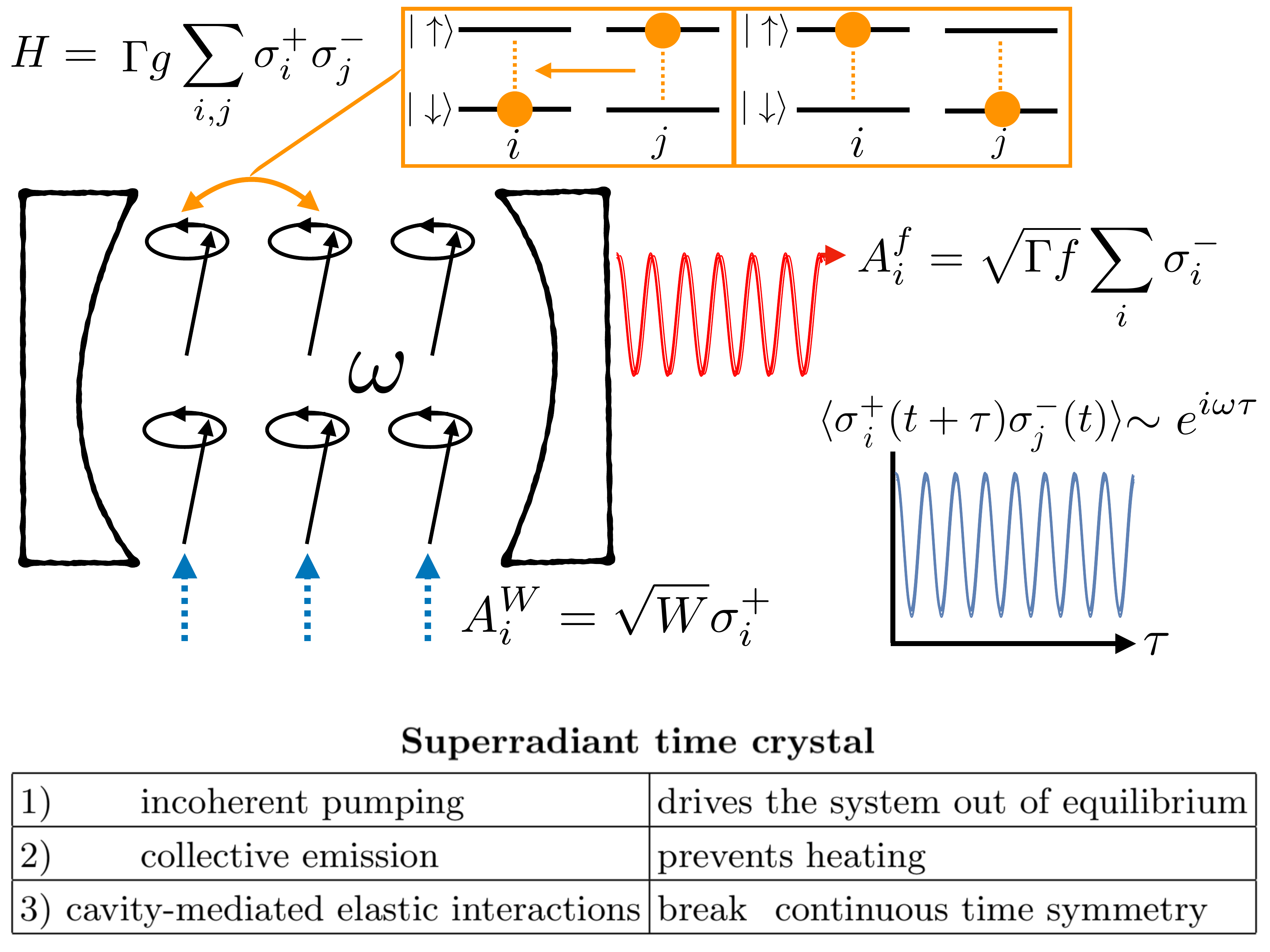}
\caption{An ensemble of $N$ spin 1/2 particles pumped at rate $W$,
experiencing  collective  emission at rate $\propto f\Gamma$ and collective spin-exchange interactions (orange),   $\propto g\Gamma$,
form the basis of the superradiant TC. The elastic interactions imprint collective spin oscillations at frequency $\omega$ spontaneously breaking the time translation
symmetry (manifested as persistent oscillations in the unequal time spin-spin correlation function in the thermodynamic limit).}
\label{fig1}
\end{figure}

In this work, we propose a paradigm shift in the concept of time crystals, which can lead to the first experimental observation of continuous time symmetry breaking, by considering an incoherently driven array of long lived dipoles
in  a cavity which are subject to  collective dissipative decay (superradiance) and elastic long range interactions (see Fig.~\ref{fig1}). Our scheme is similar in spirit to  recently
proposed quantum TCs with dissipation~\cite{Gong, Fazio,Chan,buca} and connected to the emergence of a time-periodic steady state in the
thermodynamic limit of an open quantum system. However, the key  difference is that by  applying incoherent pumping instead of a coherent, periodic drive we do not impose an external frequency which explicitly breaks  the
time translational symmetry \emph{ab initio}; rather, the latter  is only spontaneously broken by the subtle interplay between collective interactions and driving processes, and accordingly our proposal belongs  to the class of TCs which spontaneously break a continuous time translational symmetry. Moreover, the incoherent drive allows for the population
of a much larger Hilbert space compared to the fully symmetric Dicke manifold.

Within our framework, collective emission prevents unwanted heating and fulfills the role of disorder-induced localization in the Floquet TC (see Fig. 1). The balance of pumping and dissipation leads to
the stabilization of a non-equilibrium synchronized steady-state \cite{holland, thomp,zhu15} and  allows for the formation of a TC that is robust to  imperfections or environmental
disturbances in the presence of  finite but moderate elastic interactions. While the TC exists only when elastic interactions are present, we also find that if they are too strong
they can destroy the periodic order. The TC thus only exists within a finite window of interaction strengths with a width which we show grows as the square-root of the particle number. Moreover, in our system the TC emerges for any arbitrary initial state, since our driven dissipative system loses memory of the initial conditions. This is in sharp contrast with closed Hamiltonian systems, where energy conservation is always required and is always reflected in the steady-state.
The melting of the TC can be understood from the population of low-lying eigenvectors of the  Liouvillian operator that have a finite negative real component:
we find that these eigenvalues can be  linked  to  the growth of mutual information in the transient dynamics.
%

%
%
%
%

%

%
%


\begin{figure*}[t!]
        \includegraphics[width=0.3\textwidth]{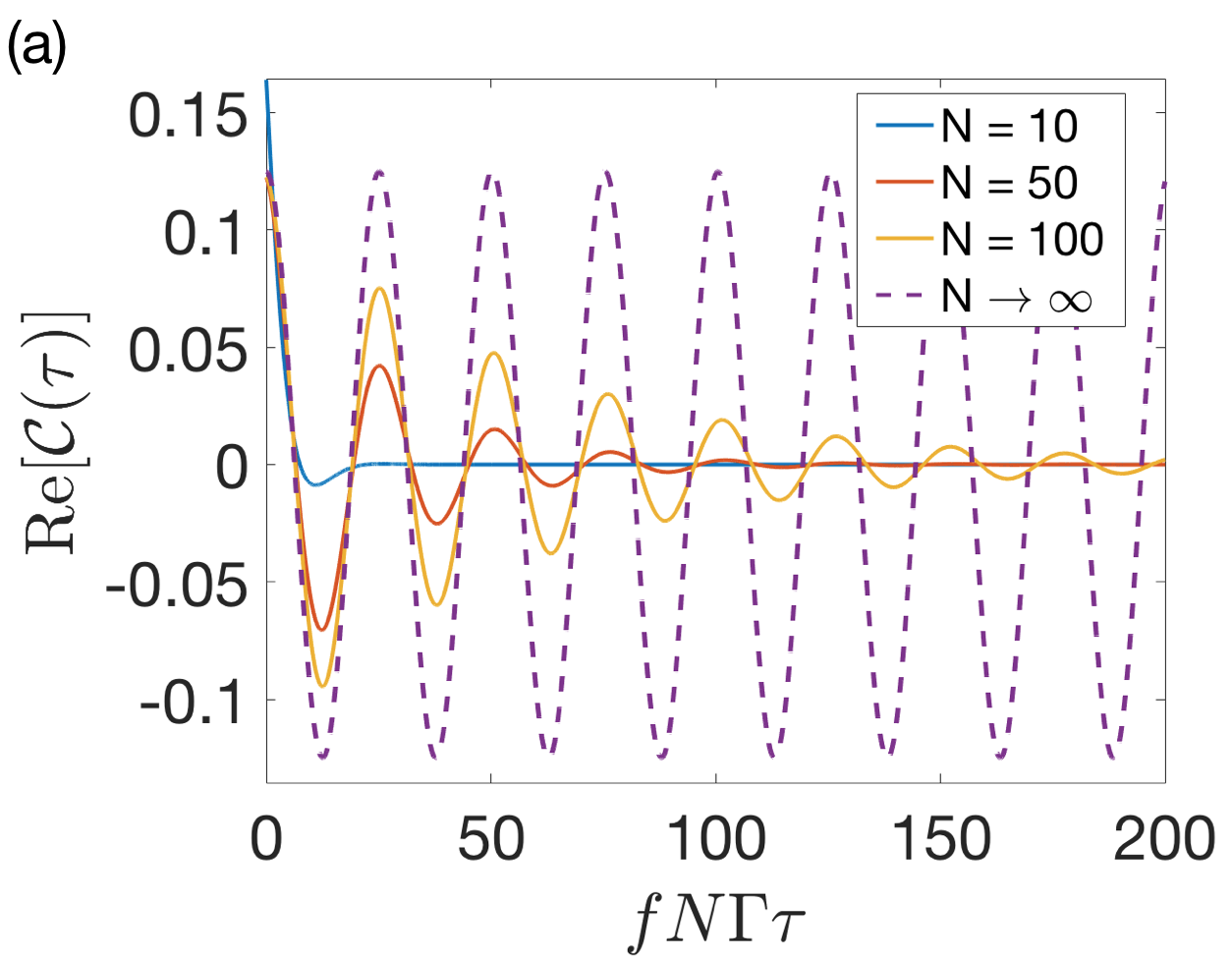}
        \includegraphics[width=0.3\textwidth]{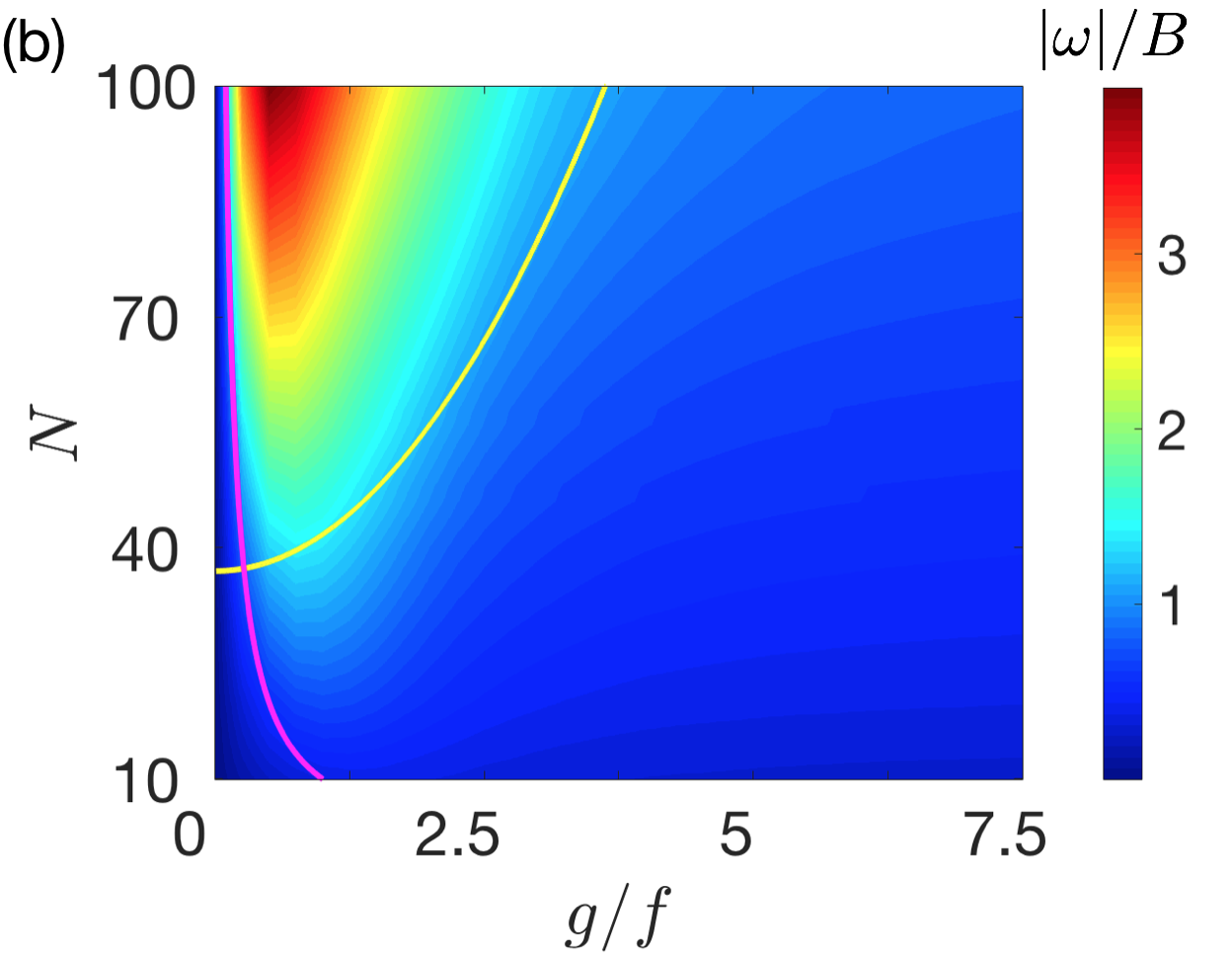}
        \includegraphics[width=0.3\textwidth]{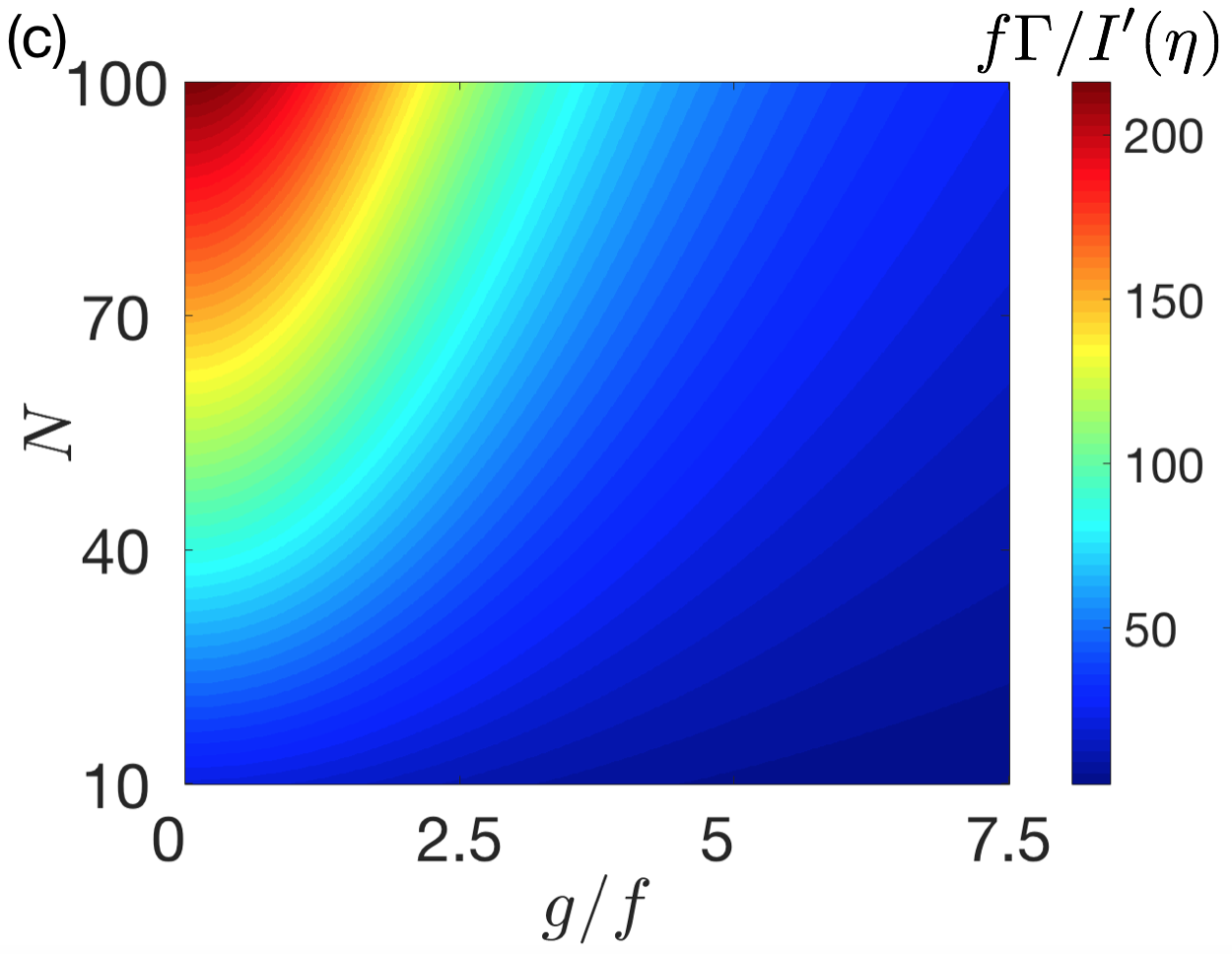}
    \caption{ (a) Real part of $\mathcal{C}(\tau)$ as a function of the characteristic time $\eta\equiv f N \Gamma \tau$ at
    optimal pumping and $g/f=1/2$ for a system of $N= 10$ (blue), $N = 50$ (red), and $N=100$ (orange) spins along with the finite size scaling prediction in the thermodynamic limit (purple).
    (b)~Extracted ratio of the absolute value of $\mathcal{C}(\tau)$ angular frequency $\vert\omega\vert$ over its spectral width $B$ (see SM \cite{supp})  vs  system size $N$ and
    interaction coupling $g/f$.  We also show a frequency contour corresponding to $\omega_{\text{MF}}^{opt}/f\Gamma \sim 5$ (purple) and a contour of mutual information growth corresponding to $f\Gamma/I'(\eta=0.03) \sim 80$ (yellow). (c) Growth rate of two-particle mutual information at short characteristic times (here we set $\eta=0.03$) starting from
    a maximally coherent array $\langle{\sigma^x_i(0)}\rangle=1$. }
    \label{fig_2}
\end{figure*}

%

\emph{Model --} We consider an ensemble of $N$ spin-$1/2$ particles, whose evolution is described by a master equation for the density matrix $\hat{\rho}$,
\begin{eqnarray}
\label{eq:dyn}
\partial_t\hat{\rho}&=&\operatorname{L}[\hat{\rho}]=-i[\hat H,\hat{\rho}]+\mathcal{L}[\hat{\rho}],\\
\hat H&=&{g \Gamma}\hat S^+\hat S^- +\sum^N_{i=1}\frac{\delta_{i}}{2}\hat\sigma^z_i,
\end{eqnarray}
where $\hat S^\pm\equiv\sum^N_{i=1}\hat\sigma^\pm_i$ and $\hat \sigma^a_i$ are the Pauli matrices ($a=x,y,z$) acting on spin  $i=1,...,N$. 
The first term in $\hat H$ describes a collective spin-exchange interaction, whilst the second describes a static  disordered  magnetic field along the $\hat{z}$-direction. Notice we avoid normalization of the interaction term by $N$ to model relevant experimental implementations of our system \cite{norcia,lerose}.
For simplicity, but without loss of generality, we assume the
$\delta_i$'s are distributed according to a Lorentzian of width $\Delta$ and zero mean.
The dissipator $\mathcal{L}[\hat{\rho}]= \mathcal{L}^W[\hat{\rho}]+ \mathcal{L}^f[\hat{\rho}]$ encodes two  channels: local,  incoherent pumping
described by  $  \mathcal{L}^W[\hat{\rho}]\equiv \sum_i \hat{A}_i^W\hat{\rho} \hat{A}_i^{W\dag}-\frac{1}{2}\{\hat{A}_i^{W\dag} \hat{A}_i^W,\hat{\rho}\}$ with
$\hat{A}_i^{W}=\sqrt{W}\sigma^+_i$, and collective emission  described by $\mathcal{L}^f[\hat{\rho}]\equiv  \hat{A}^f \hat{\rho} \hat{A}^{f\dag}-\frac{1}{2}\{ \hat{A}^{f\dag}   \hat{A}^f,\hat{\rho}\}$
with $ \hat{A}^f=\sqrt{f \Gamma}\hat{S}^-$. 
The parameter $\Gamma$ sets the scale of the  spin-spin interactions, while $g$ and $f$ are dimensionless parameters characterizing the  relative strength of their corresponding elastic and dissipative part respectively. The use of a master equation \cite{Lax,Car} to deal with the dissipative processes  is extremely accurate for the experimental systems discussed below \cite{norcia,Shankar2017}.
The incoherent nature of the pumping preserves the $U(1)$ phase  symmetry of the dynamics (which can be seen by the invariance of $\mathcal{L}[\hat{\rho}]$  under
the transformation $\hat \sigma^+_j\to \hat \sigma^+_j e^{i\phi}$). In fact, in the
steady state the condition   $\langle \hat \sigma^+_j(t\to \infty)\rangle =0$, is always satisfied. Moreover, the incoherent pumping allows coupling between states with
different total $S$, with $S(S+1)$ the eigenvalues of the $\vec{\hat S}\cdot \vec{\hat S}$ operator, and thus the dynamics is not restricted to the collective $S=N/2$ manifold.

%
%
%

\emph{Mean-field analysis --} We start with a simple mean-field analysis which illustrates how synchronization emerges in a dissipative setting. It assumes
that the many-body density matrix  can be written as a tensor product of single-spin density matrices and thus neglects spin-spin correlations. This is equivalent to a description of the system in
terms of $N$ Bloch vectors, $\vec{\mathcal{S}}_i=\{R_i\cos\phi_i,\,R_i\sin\phi_i,s_i\}$, where $R_i$ represents the radius of the Bloch vector projected into the $xy$-plane, $\phi_i$ is the azimuthal angle measured from the positive $x$-axis, $s_i$ is the $z$-component, and $i=1,...,N$. In the mean-field treatment the effect of the elastic and dissipative interactions generates a
 self-adjusting  effective complex magnetic field identically experienced by each spin in the ensemble due to interactions with the other spins. The corresponding non-linear Bloch equations are presented
 in the SM \cite{supp}. Here, we focus on  the dynamics of the azimuthal phases,
\begin{equation}
\label{eq:MF}
\frac{d\phi_i}{dt} =  \delta_i + \frac{\Gamma s_i}{R_i} \sum_{j\neq i}R_j \left[ f\sin(\delta \phi_{ij}) - 2g\cos( \delta \phi_{ij})\right],
\end{equation}
with $\delta \phi_{ij}=\phi_j - \phi_i$. From direct inspection, we see that in Eq.~\eqref{eq:MF} the term proportional
to $f$ can be identified with a similar term in the Kuramoto model~\cite{kuramoto1}, the iconic model used to  describe the emergence of phase synchronization in classical non-linear oscillators.
For synchronization to occur, the coupling strength per oscillator, here proportional to  $f \Gamma s_i>0$ , must be positive and large enough to compensate for the dephasing generated by the different single particle frequencies. This condition is only possible in the presence of incoherent pumping and thus intrinsic to our setup since a coherent drive does not lead to population inversion in the steady-state \cite{Car2, Drumm}. The term proportional to $g$, arising from the real part of the effective magnetic field, also present in the Kuramoto-Sakaguchi model \cite{kuramoto2},  is responsible for imprinting a  collective spin rotation on the non-equilibrium steady state. The effective field not only induces a net collective precession but also favors spin alignment and  self-rephasing  against the depolarization induced by the inhomogeneous field as theoretically and experimentally demonstrated in prior work~\cite{du,rey,deutsch,kleine,martin,norcia,Bromley}.
Therefore, both $f$ and $g$ are responsible for the rigidity of the time crystal.

Diagnosis of the TC regime at the mean-field level proceeds by
assuming the existence of a synchronized solution of the mean-field  equations, namely
setting $\delta_i$ to zero and determining self-consistently the associated frequency which governs the collective oscillatory dynamics. Later we will  restore the detunings and show robustness of the TC to those
imperfections. We  define the normalized collective order parameter $S^+(t)\equiv\sum_{j=1}^N R_je^{i\phi_j}$, and  assume the following scaling
form $S^+(t)/N=Ze^{i\omega_{\text{MF}} t}$, looking for a solution of the equations of motion which is stationary in the frame co-rotating at the angular
frequency $\omega_{\text{MF}}$. This mean-field solution can feature collective oscillations in  the order parameter $S^+(t)$  breaking  the $U(1)$-symmetry of
the microscopic dynamics (Eq.~\eqref{eq:dyn}). This is not the case in the exact solution that must preserve the  $U(1)$-symmetry, $\langle \hat S^+(t\to \infty)\rangle =0$, where the long time limit is taken for a fixed $N$.

The system  synchronizes when $Z$ acquires a positive real value, which
is self-consistently determined from the system's parameters \cite{holland,zhu15}. For $W< f \Gamma$ the mean-field equations do not admit a synchronized solution for $Z$, as single particle emission dominates the dynamics
and the system depolarizes completely. The same occurs if $W> f N \Gamma$, since in this regime pumping dominates, heating the system into a trivial incoherent state fully polarized along the $\hat{z}$-direction;
however, for $N>8$ there always exists a finite window of pumping parameters in which the system enters a synchronized phase, with non-vanishing order parameter featuring collective,
synchronous, oscillatory dynamics at the angular frequency $\omega_{\text{MF}}$. There exists an optimal pumping rate, $W_{opt}= f N \Gamma/2$ (assuming $N\gg1$), for which
the magnitude of the order parameter reaches a maximum value, $Z\simeq 1/\sqrt{8}$ (see SM \cite{supp}). While $Z$ is found to be independent of $g$, $\omega_{\text{MF}}$ is proportional to $g$ and approaches $\omega_{\text{MF}}^{opt}=g N \Gamma /2$
at optimal pumping. The rigidity of this frequency will be discussed below.

\emph{ Quantum model --}
We  now extend our understanding of the formation of the TC beyond the mean-field approximation. Specifically, we study the order parameter for the TC given in terms of the two-time
correlation function,
\begin{equation}
\mathcal{C}(\tau)\equiv \lim_{t \rightarrow \infty} \frac{\sum_{i=1,j=1} ^{N} \langle{\sigma_i}^+(t+\tau){\sigma_{j}}^-(t)\rangle}{N^2}. \label{twopoint}
\end{equation} Our analysis is based on an efficient exact numerical solution of the master equation [Eq.~\eqref{eq:dyn}]~that uses the spin permutation symmetry to drastically
improve the exponential scaling of the Liouville space from $4^N$ to $\mathcal{O}(N^3)$ \cite{Garttner,Xu,Shammah}.  $\mathcal{C}(\tau)$ is computed via the linear quantum regression theorem~\cite{Car}, $
\langle \hat \sigma_i^+(t+\tau)\hat \sigma^-_j(t)\rangle=\operatorname{Tr}\left[{\hat \sigma_i^+e^{\operatorname{L}\tau}}[\hat \sigma^-_j\hat{\rho}(t)]\right]$, which is exact for the case of a master equation.

The square root of the equal time correlator, $\sqrt{\mathcal{C}(0)}\equiv Z_Q$, corresponds to the quantum analog of $Z$.
In close agreement with  the mean-field solution, $Z_Q$ is found to be nonzero within a window of pumping $W$ where the system synchronizes, and reaches a  maximum value, $Z_Q ^{opt} \approx 1/\sqrt{8}$,  almost independently of $g$ at $W_{opt}$, when the system is maximally synchronized (see SM \cite{supp}). On the other hand  $\mathcal{C}(\tau>0)$ is highly dependent on $g$. At optimal pumping, large  $N$ and
finite but moderate interactions, $1/N\lesssim g/f\lesssim  \sqrt{N}$, the order parameter $\mathcal{C}(\tau)$ oscillates at the mean-field angular frequency  $\omega_{\text{MF}}^{opt}$. The oscillations  slowly decay but  appear to become persistent  in the thermodynamic limit, thus signaling  the emergence of a time crystal. In other words, in units of the TC periodicity, the decay time goes to infinity in the thermodynamic limit.
Figure \ref{fig_2}~(a) illustrates this aspect, where  $\mathcal{C}(\tau)$ is plotted as a function of the characteristic time $\eta\equiv f N \Gamma \tau$  for $N = 10,$ $50$ and $N=100$.

\begin{figure*}[t!]
        \includegraphics[width=0.3\textwidth]{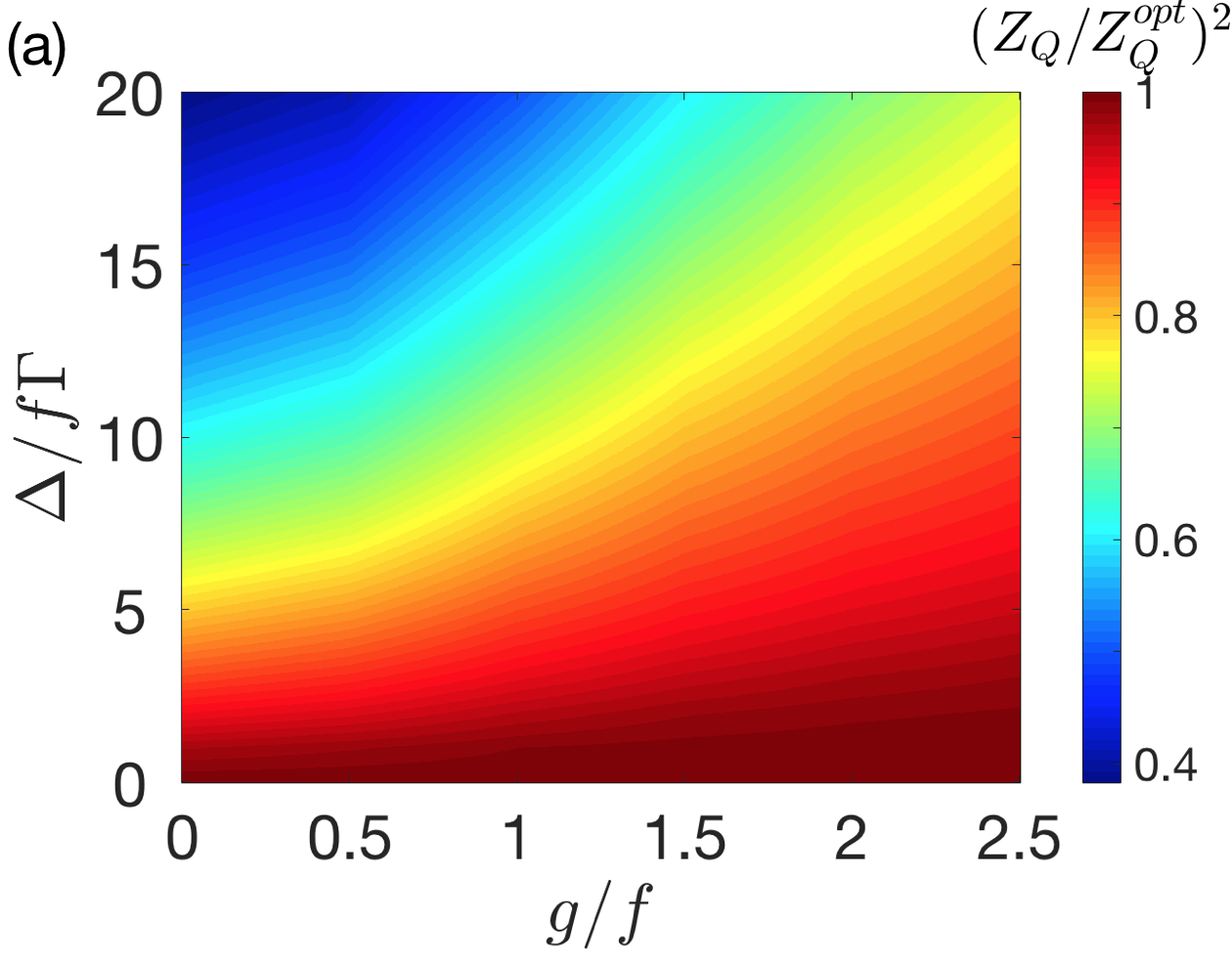}
        \includegraphics[width=0.3\textwidth]{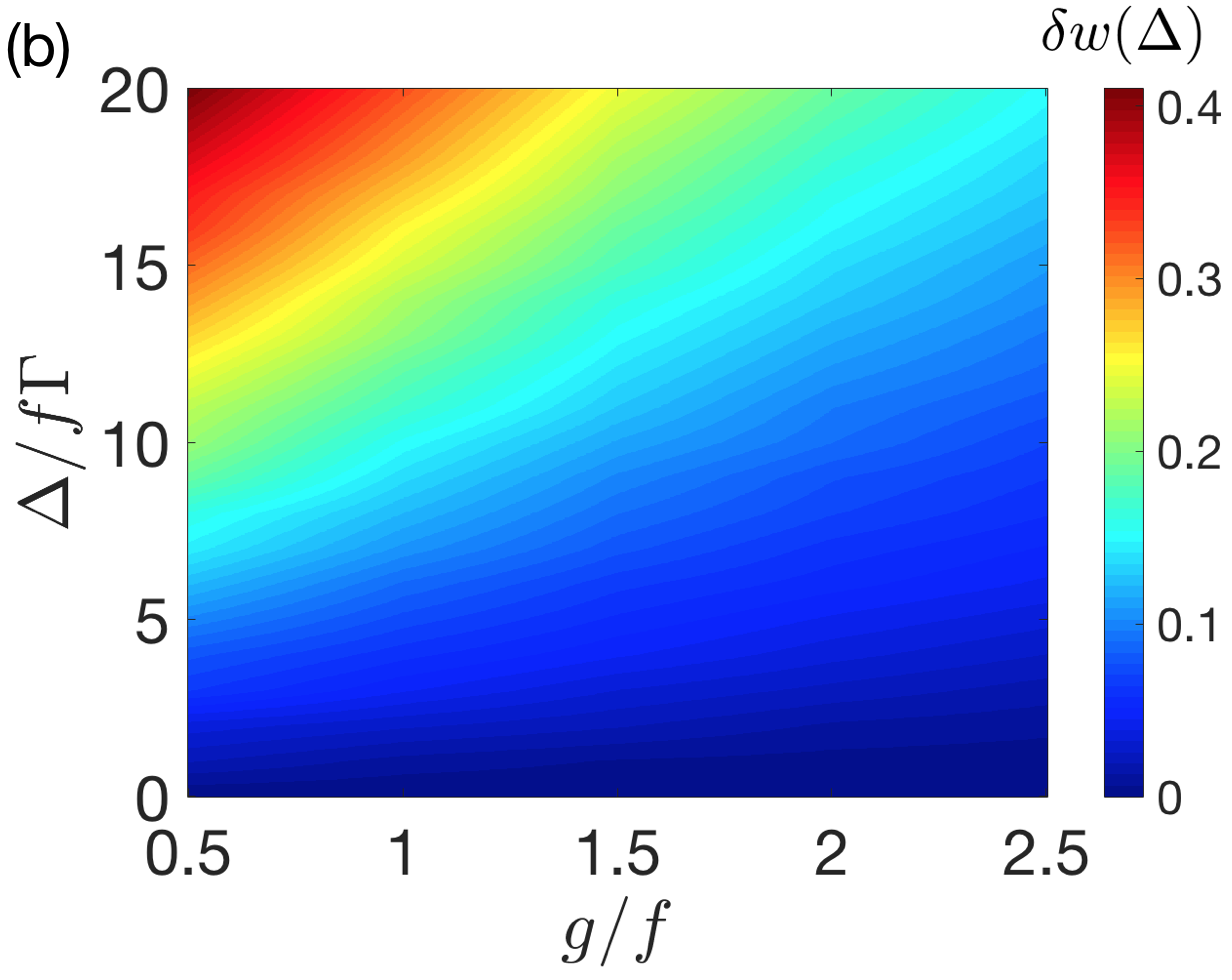}
        \includegraphics[width=0.3\textwidth]{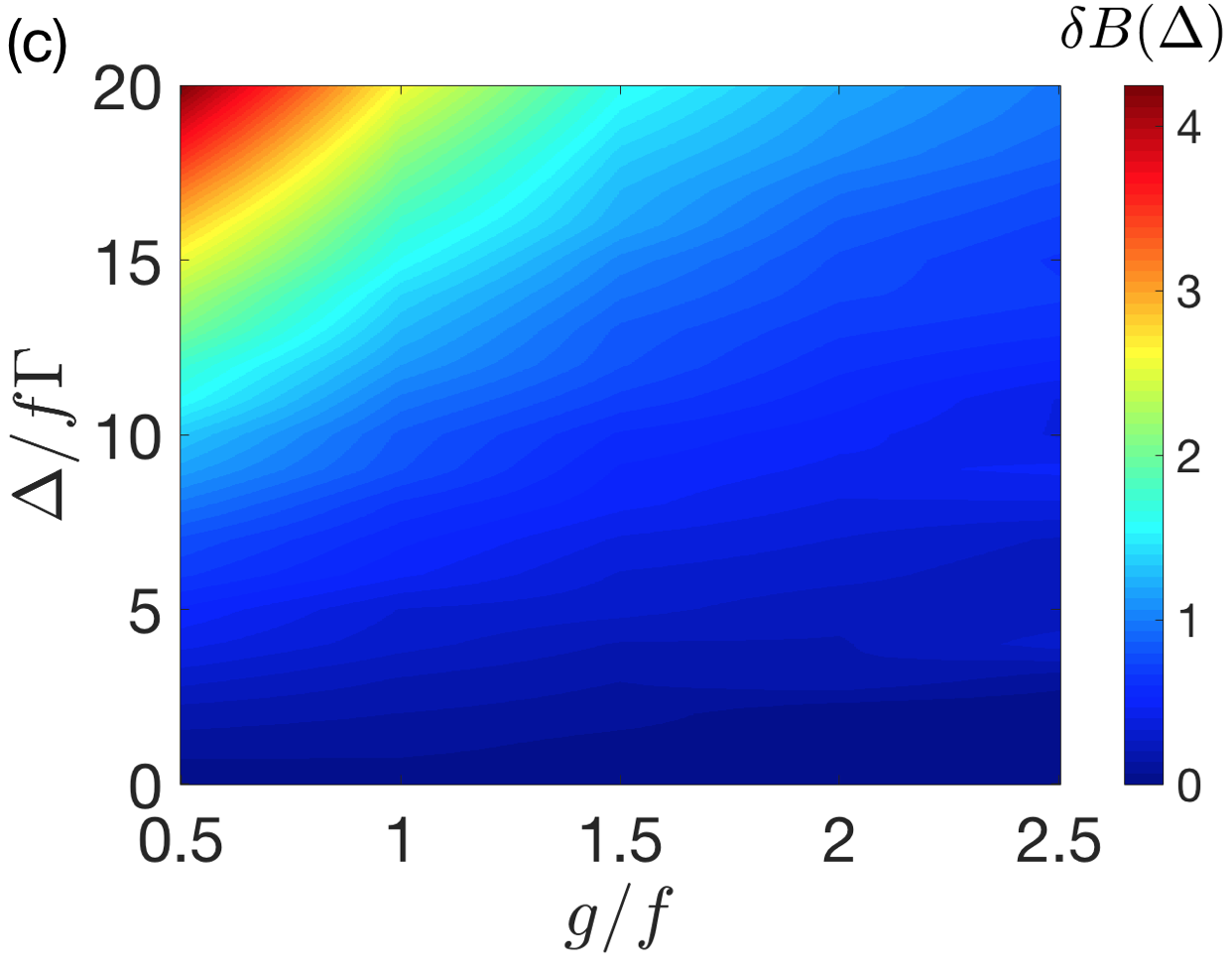}
    \caption{ Finite interactions  protect the TC  against weak  disorder as can be seen in: (a) The robustness of the  averaged  $\mathcal{C}(0)$, the insensitivity of the  time crystal's (b)   frequency, $\delta\omega(\Delta)$, and (c)  spectral width, $\delta B(\Delta)$,  to  weak disorder $\Delta/(f \Gamma)$ for the relevant window of interactions $g/f$. All plots are computed using the second order cumulant expansion at optimal pumping  $W_{opt}$ for $N=100$ spins. }
    \label{fig_3}
\end{figure*}


Large many-body correlations  can lead to melting of the time crystal~\cite{Yao,Khe}.
For the purposes of quantifying the window of stability of the  time crystal, we use  the absolute value of the ratio between $\mathcal{C}(\tau)$  oscillation angular frequency, $\omega$, and  corresponding  decay rate, $B$ (or bandwidth)
which is  portrayed  in Fig.~\ref{fig_2}(b). 
In our open system we can understand this behavior directly
from the quantum regression theorem. While  the lowest energy eigenstates of the Liouvillian with pure imaginary eigenvalue are the ones that determine the oscillatory character in the late-time synchronized state, 
excited eigenstates  can contribute to the dynamics of the unequal time  correlator through the term $e^{\operatorname{L}\tau}$. For moderate interactions, low-lying Liouvillian  eigenvectors will have eigenvalues with
nonzero imaginary part and small, negative real part. These eigenvalues are dominant and determine the oscillation frequency. As interaction strength increases, however, so does the magnitude of
the real part, which results in damping of the TC \cite{supp}. Such mechanism is also responsible for the growth of many-body correlations during the transient dynamics and thus directly manifest in the mutual
information defined as  $I_{AB}\equiv S_A+S_B-S_{AB}$, where $S_{\alpha}=-\operatorname{Tr}\Big[ \hat \rho_{\alpha} \log[\hat\rho_{\alpha}]\Big]$ is the Von Neumann entropy computed from the reduced
density matrix $\hat \rho_{\alpha} $ of the subsystem $\alpha= A, B, AB$ (${AB}$ is the joint subsystem).

To establish a more formal connection between the TC stability and the growth of many-body correlations, we have computed the derivative of the mutual information  for the case when  $A$ and $B$ are  single spin
subsystems, starting    from a maximally coherent initial state (all spins pointing along the $\hat{x}$-direction).  We find that in the large $N$ limit and short characteristic times $\eta\ll 1$ (respect to the oscillation period), with   $\eta\equiv f N \Gamma \tau$, the growth  rate of
the mutual information approaches  $d I_{AB}/d\tau \sim \frac{f \Gamma}{2}\big(\frac{1}{N}+\frac{4g^2}{ N f^2}\eta\big)$ and thus remains irrelevant for $g/f\lesssim  \sqrt{N}$.
This  parameter regime is consistent with the range of $g$ values where we observe that the time
crystal forms. Outside this region, $I_{AB}$ grows rapidly with increasing $g/f$ (see Fig.~\ref{fig_2}c). We explicitly indicate the contours $g/f \propto 1/N$ and  $g/f \propto \sqrt{N}$ set by $f\Gamma/\omega_{\text{MF}}^{opt}$ and the onset of fast mutual information growth, respectively, in Fig.~\ref{fig_2}(b).
From these considerations we can conclude that the superradiant crystal only exists in the parameter regime where many-body correlations are subdominant and thus
it can be regarded as an emergent semi-classical non-equilibrium state of matter.

\emph{Robustness to disorder --}  We now  investigate  the impact of inhomogeneous dephasing, $\delta_i\neq0$. For this, we resort to  a second-order cumulant expansion~\cite{bonitz}, since in the absence of
permutation symmetry exact numerical calculations are constrained to small systems, $N\lesssim 15$. 
The complete set of equations of motion  and equations for the two-time correlations are presented in the SM \cite{supp}.
We find excellent agreement between the  cumulant expansion and  exact numerical solution of $\mathcal C(\tau)$  in the region where a stable TC is expected  for the homogeneous ($\delta_i = 0$) case,
as well as for small system sizes in the presence of disorder \cite{supp}.

In Fig.~\ref{fig_3}(a) we show  the robustness of the  averaged    equal time  correlator $Z_Q=\sqrt{ \mathcal{C}(0)}$ evaluated at optimal pumping $W_{opt}$ to weak disorder $\Delta/(f \Gamma)$ within the
relevant window of interactions $g/f$.
One observes that finite elastic interactions protect the synchronized state against disorder, preserve  phase coherence and  favor spin alignment. While  similar phase locking effects in the transient dynamics
have been experimentally reported  in cold atom experiments \cite{du,rey,deutsch,kleine,martin,norcia,Bromley}, the interesting feature observed here is that  the phase locking is achieved in the steady
state of a driven dissipative system. Panels (b) and (c) of Fig.~\ref{fig_3} portray the  variation in oscillation frequency, $\delta\omega (\Delta)\equiv \left[\omega(\Delta)-\omega(0)\right]/\omega(0)$, and in the
spectral width, $\delta B (\Delta)\equiv \left[B(\Delta)-B(0)\right]/B(0)$, of the averaged two-time correlation function determined from  the cumulant expansion. The observed rigidity of the
frequency  also agrees with the simpler mean-field predictions \cite{supp} which allow us to derive an analytic expression for the protection in the weak
disorder limit: $\delta\omega (\Delta)\sim \frac{\sqrt{8}\Delta}{N \Gamma \sqrt{ f^2+ 2 g^2}}$,  where we observe the  $1/N$ suppression gained from the collective nature of the elastic and dissipative interactions.

\emph{ Experimental Realization and Outlook--} The superradiant crystal can be directly realized using an  array of incoherently pumped  atomic dipoles tightly trapped by a deep
optical lattice that is supported by an optical cavity. The cavity couples two relevant internal states  of the atoms, and operates in the  bad cavity  limit 
where the bare atomic linewidth $\gamma$ is significantly smaller than the cavity linewidth $\kappa$. In this regime  the cavity  photons do not participate actively in the dynamics  but instead mediate
collective dissipative decay (superradiant emission)~\cite{Haroche,holland, thomp}, with $f\Gamma \propto \kappa/(4\delta_c^2 + \kappa^2)$, and   elastic exchange interactions, with $g\Gamma \propto \delta_c/(4\delta_c^2 + \kappa^2)$, which can be independently controlled by varying  the cavity  detunning  $\delta_c$ from the atomic transition. The signature of the TC
can then be directly observed in the spectrum of the light leaked from the cavity~\cite{norcia}. A similar implementation can be realized by replacing the cavity photons by phonons in an ion crystal \cite{Shankar2017}. In the case of the cavity, the order of magnitude for $f\Gamma/2\pi$ and $g\Gamma/2\pi$ is approximately $10^{-4}$ Hz.  For typical atom number in the cavity, $N \approx 10^5$, the  TC oscillation frequency  approaches $\approx 10$ Hz. In the case of the ion crystal, we have $f\Gamma/2\pi = g\Gamma/2\pi \approx 6$ Hz.  In this case the  TC oscillation frequency for typical ion  number $N \approx 10^2$  approaches $\approx 10^3$ Hz.

Having demonstrated the rigidity of the TC to dephasing, now we discuss its rigidity to variations in the system's parameters.
For the proposed implementation, $\omega^{opt}_{\text{MF}}\propto N   \delta_c/( 4\delta_c^2+\kappa^2)$.  From this expression, one can see $\omega^{opt}_{\text{MF}}$
is not highly  sensitive to variations in  the cavity linewidth, $\kappa$, but on the contrary  it is linearly sensitive to  variations on  $\delta_c$ and $N$.
Systematics in the  cavity detuning, nevertheless,  can be currently  controlled at the subhertz level by locking the cavity to a state-of-the-art clock laser \cite{Bloom2014}.
Fluctuations in $N$ can be also  suppressed by operating the system in a three dimensional  optical lattice in the band or Mott insulator regimes \cite{Campbell2017} and spectroscopically selecting a fixed region of the atomic array \cite{Marti2018}.

In summary, we have proposed and investigated the emergence of a TC in a  many-body driven dissipative quantum system.
By investigating its stability to quantum correlations we showed that it only exists in the parameter regime where many-body correlations are subdominant and thus  can be
regarded as an emergent semi-classical non-equilibrium state of matter. However, it is important to emphasize that this system  is fundamentally  distinct from the prototypical laser. This can be seen from the fact that the working mechanism of a laser is stimulated emission, an ingredient absent in our setup. Lasing action is possible even in a single atom system or in the absence of coupling between the atomic dipoles. The superradiant TC, on the contrary, is a genuine many-body phenomenon that happens in the bad cavity limit where the mean photon number in the cavity is less than one. However, even without  stimulated emission, superradiance can happen due to collective interactions in a many-body array of long lived atomic dipoles. The superradiant TC is thus  a genuine
many-body phenomenon which  can produce spectrally pure light  and
might find direct applications in ``quantum-interaction enhanced'' sensing.




\emph{Acknowledgements ---}
We acknowledge useful discussions with I. Bloch, M. Foss-Feig, M. Holland, A. Shankar, J. Thompson and P. Zoller. The authors acknowledge support from Defense Advanced Research Projects Agency (DARPA) and Army Research Office grant W911NF-16-1-0576, NSF grant PHY1820885, JILA-NSF grant PFC-173400, and the
Air Force Office of Scientific Research  FA9550-18-1-0319 and its Multidisciplinary University Research Initiative grant FA9550-13-1-0086, and NIST. JM is supported by the European Union's Framework Programme for Research and Innovation Horizon 2020 2014-2020) under the Marie Sklodowska-Curie Grant Agreement No. 745608~('QUAKE4PRELIMAT'). B. Zhu was supported by the NSF through a grant for the Institute for Theoretical Atomic, Molecular, and Optical Physics at Harvard University and the Smithsonian Astrophysical Observatory.

\bibliography{biblio_time_crystal}
\end{document}